\documentclass[apjl]{emulateapj}
\usepackage{epsfig}
\usepackage{amsmath,amssymb,bm}
\usepackage{color}
\usepackage[varg]{txfonts}

\begin{document}
\label{firstpage}

\title[Kozai--Lidov cycles in fluid disks]{The Kozai--Lidov Mechanism in Hydrodynamical Disks}

\author{Rebecca G. Martin\altaffilmark{1,6}} 
\author{Chris Nixon\altaffilmark{1,7}} 
\author{Stephen H. Lubow\altaffilmark{2}}
\author{Philip J. Armitage\altaffilmark{1}} 
\author{Daniel J. Price\altaffilmark{3}} 
\author{Suzan~Do\u{g}an\altaffilmark{4,5}}
\author{Andrew~King\altaffilmark{5}}
\affil{\altaffilmark{1}JILA, University of
  Colorado \& NIST, UCB 440, Boulder, CO 80309, USA}
\affil{\altaffilmark{2}Space Telescope Science Institute, 3700 San
  Martin Drive, Baltimore, MD 21218, USA }
\affil{\altaffilmark{3}Monash Centre for Astrophysics (MoCA), School
  of Mathematical Sciences, Monash University, Vic. 3800, Australia}
\affil{\altaffilmark{4}University of Ege, Department of Astronomy \& Space Sciences, 
Bornova, 35100, ${\dot {\rm I}}$zmir, Turkey}
\affil{\altaffilmark{5}Department of Physics and Astronomy, University of Leicester, 
University Road, Leicester LE1 7RH, UK}
\affil{\altaffilmark{6}Sagan Fellow} \affil{\altaffilmark{7}Einstein
  Fellow}

\begin{abstract}
We use three dimensional hydrodynamical simulations to show that a
highly misaligned accretion disk around one component of a binary
system can exhibit global Kozai--Lidov cycles, where the inclination
and eccentricity of the disk are interchanged periodically.  This has
important implications for accreting systems on all scales, for
example, the formation of planets and satellites in circumstellar and
circumplanetary disks, outbursts in X-ray binary systems and accretion
on to supermassive black holes.
\end{abstract}

\section{Introduction}
\label{intro}

Disks that orbit objects in a variety of types of astrophysical binary
systems can sometimes be misaligned with respect to their binary
orbital planes.  When material that is misaligned to the binary orbit
is accreted, a misaligned accretion disk can form around the primary
and/or the secondary masses. Supermassive black hole binaries are
likely to accrete material in a chaotic fashion \citep{KP2006,KP2007}
and therefore highly inclined disks around each black hole are
expected \citep{Nixonetal2013,KN2013}. Also, warped disks have been
observed in AGN maser disks \citep[e.g.][]{Caproni2006,Martin2008}.

If a young binary star system accretes material after its formation
process, the material is likely to be misaligned to the binary orbit
and so misaligned disks may form around young stars and become the
sites of planet formation \citep{Bateetal2010}.  For widely separated
stars in a binary, greater than 40 AU, a misaligned disk may occur
because the stellar equatorial inclinations, based on spins, are
observationally inferred to be misaligned with respect to the binary
orbital planes \citep{Hale1994}.  More speculatively, highly
misaligned disks could form around young giant planets if the planet's
inclination to the protoplanetary disk is sufficiently high.

Evolved binary star systems such as low mass X-ray binaries
\citep{NixonandSalvessen2014}, microquasars
\citep[e.g][]{Maccarone2002,MartinReisPringle2008,Martinetal2008} and
Be/X-ray binaries \citep{Martinetal2011} are thought to have disks
that are misaligned to the binary orbit. When a star in a binary star
system undergoes an asymmetric supernova explosion, the explosion can
leave the spin of the unexploded star in highly misaligned state with
respect to the binary orbit. In this configuration, a misaligned disk
may form from material ejected by the unexploded star
\citep{Martinetal2009b,Martinetal2010}.

Kozai--Lidov (KL) oscillations occur in highly misaligned test
particle orbits around one component of a binary, where the particle's
inclination is periodically exchanged for eccentricity
\citep{Kozai1962,Lidov1962}. During this process, the component of the
angular momentum that is perpendicular to the binary orbital plane is
conserved, which is expressed as
\begin{equation}
 \sqrt{1-e_{\rm p}^2}\cos i_{\rm p} \approx {\rm const},
\label{hz}
\end{equation}
where $i_{\rm p}$ is the inclination of the particle orbital plane
relative to the binary orbit plane and $e_{\rm p}$ is the eccentricity
of the test particle.  A test particle that is initially on a circular
and highly misaligned orbit undergoes oscillations of its orbital
plane involving closer alignment (higher values of $|\cos i_{\rm p}|$)
and therefore higher values of its eccentricity $e_{\rm p}$. For these
oscillations to occur, the initial inclination of the test particle
orbit $i_{\rm p0}$ must satisfy the condition that $\cos^2{i_{\rm p0}}
< \cos^2{i_{\rm cr}} = 3/5$. This condition requires that $ 39^\circ
\la i_{\rm p0} \la141^\circ$.  Furthermore, inclination values $i_{\rm
  p}$ during the oscillations are bounded by the condition that
$\cos^2{i_{\rm p0}} \le \cos^2{i_{\rm p}} \le \cos^2{i_{\rm cr}}$. As
follows from equation (\ref{hz}), the maximum eccentricity that an
initially circular particle orbit can achieve is given by
\begin{equation}
e_{\rm max}=\sqrt{1-\frac{5}{3}\cos^2 i_{\rm p0}}
\label{emax}
\end{equation}
\citep[e.g.][]{Innanen1997}.      

The KL mechanism for orbiting objects has been extensively studied in
the literature.  For example, it can occur for asteroids
\citep{Kozai1962}, artificial satellites \citep{Lidov1962}, triple
star systems \citep{Eggleton2001, Fabrycky2007}, planet formation with
inclined stellar companions \citep{Wu2003, Takeda2005}, inclined
planetary companions \citep{Nagasawa2008}, merging supermassive black
holes \citep{Blaesetal2002}, stellar compact objects
\citep{Thompson2011} and blue straggler stars \citep{Perets2009}.
\cite{Batygin2011} and \cite{Batygin2012} investigated the evolution
of self-gravitating, but pressureless and inviscid, misaligned disks
in binary systems. However, no KL oscillations were found.  Recently,
\cite{Teyssandier2013} showed that an external gas disk can induce
Kozai oscillations in the orbit of a misaligned companion. However, to
our knowledge, the effect of the KL mechanism acting {\it on} a
hydrodynamical disk has not yet been investigated.  In this work we
explore this process with three dimensional hydrodynamical simulations
of misaligned disks in binary systems.

\section{Test Particle Orbits}

We first consider ballistic particle orbits around the primary of a
circular orbit binary system. The primary has mass $M_1$, the
secondary has mass $M_2$ and they orbit at a separation $a$.  The
total mass of the binary is $M=M_1+M_2$.

\begin{figure}
\includegraphics[width=8.4cm]{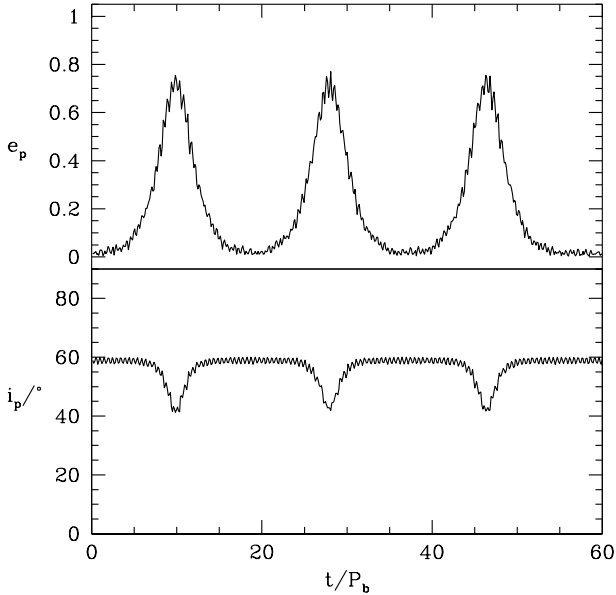}
\caption{The eccentricity and inclination evolution of a test particle
  around the primary of an equal mass binary. The particle is
  initially at a radius $d=0.2\,a$ in a circular orbit from the
  primary mass at an inclination $i_{\rm p0}=60^\circ$.}  
\label{orbits}
\end{figure}

In Fig.~\ref{orbits} we show the eccentricity and inclination
evolution of a particle that has an initial distance of $d=0.2\, a$
from the primary and inclination $i_{\rm p0}=60^\circ$ to the binary
orbital plane.  The maximum eccentricity reached is similar to that
predicted by equation~(\ref{emax}), $e_{\rm max}\approx 0.76$.  The
analytic period for KL cycles is
\begin{equation}
\frac{\tau_{\rm KL}}{P_{\rm b}} \approx \frac{M_1+M_2}{M_2}\frac{P_{\rm b}}{P_{\rm p}}(1-e_{\rm b}^2)^\frac{3}{2}
\label{tauKL}
\end{equation}
\citep[e.g.][]{Kiseleva1998}, where the orbital period of the binary
is $P_{\rm b}=2\pi/\Omega_{\rm b}$, the eccentricity of the binary is
$e_{\rm b}$ (in our circular binary case, $e_{\rm b}=0$) and the
orbital period of the particle about the primary is $P_{\rm
  p}=2\pi/\Omega_{\rm p}$. However, we find in our test particle
simulations that the timescale $\tau_{\rm KL}$ depends on the initial
inclination of the test particle. This dependence is not included in
this formula.  We determined that equation~(\ref{tauKL}) is valid up
to a factor of a few, due to this inclination dependence.  With
$M_2/M_1=1$ and $d=0.2\,a$, we find in equation~(\ref{tauKL}) that
$\tau_{\rm KL}=15.8\,P_{\rm b}$, similar to that displayed in the
particle orbits in Fig.~\ref{orbits}.  We find that the oscillatory
behaviour occurs only for inclinations $i \gtrsim 40^\circ$, in line
with the KL mechanism.

In the next Section we investigate the response of a hydrodynamical
(pressure and viscous internal forces) disk that satisfies the
criteria for the KL oscillations to occur on a test particle. As far
as we know, this is the first time that this has been investigated.

\section{Hydrodynamical Disk Simulations}
\label{sim}

\begin{table}
\caption{Parameters of the initial disk set up for a circular equal
  mass binary with total mass, $M$, and separation, $a$.} \centering
\begin{tabular}{lllll}
\hline
Binary and Disk Parameters & Symbol & Value \\
\hline
\hline
Mass of each binary component &  $M_1/M = M_2/M$ & 0.5 \\
Accretion radius of the masses & $R_{\rm acc}/a$    & 0.025  \\
Initial  disk mass & $M_{\rm di}/M$ & 0.001 \\
Initial disk inner radius & $R_{\rm in}/a$ & 0.025 \\
Initial disk outer radius & $R_{\rm out}/a$ & 0.25 \\
Disk viscosity parameter & $\alpha$ & $0.1-0.12$ \\
Disk aspect ratio & $H/R (R=R_{\rm in})$ & 0.035 \\
   & $H/R (R=R_{\rm out})$ & 0.02 \\
Initial disk inclination & $i$ & $60^\circ$ \\ 
\hline

\end{tabular}
\label{tab}

\end{table}

\begin{figure*}
\begin{center}
\includegraphics[width=17cm]{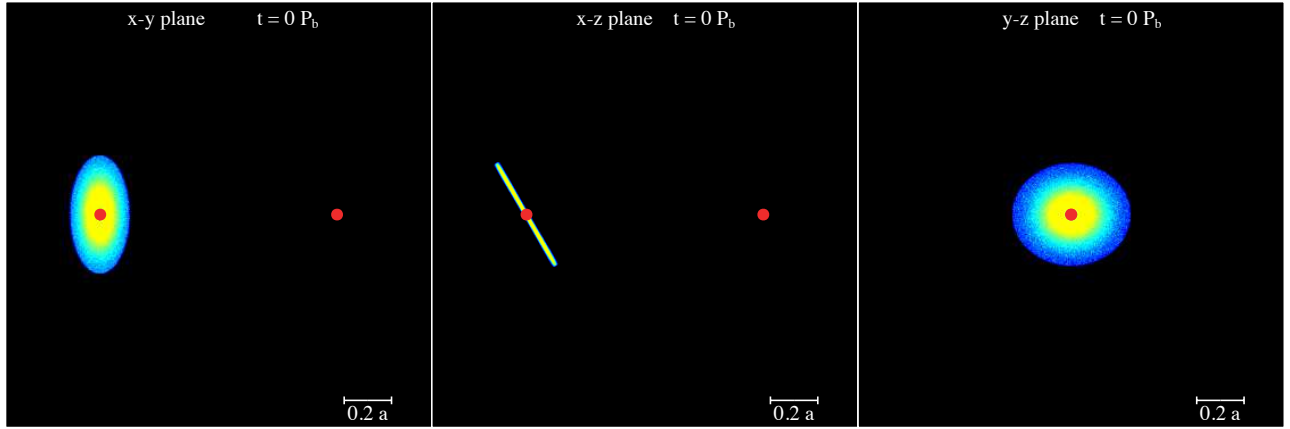}
\end{center}
\caption{The initial disk set up for the SPH simulation of a binary
  (shown by the red circles) with a disk around the primary mass. The
  size of the circles denotes the SPH accretion radius. The colour of the
  gas denotes the column density with yellow being about two orders of
  magnitude larger than blue.  The left panel shows the view looking
  down on to the $x$-$y$ binary orbital plane and the middle and right
  panels show the views in the binary orbital plane, the $x$-$z$ and
  $y$-$z$ planes, respectively.  Initially the disk is circular and
  flat, but tilted from the binary orbital plane by $60^\circ$. Note
  that in the right hand panel the primary star and the secondary star
  overlap with each other in projection. }
\label{initial}
\end{figure*}

In this Section we consider the evolution of a highly misaligned fluid
disk, around one component of a circular equal mass binary.  We use
the smoothed particle hydrodynamics (SPH; e.g. \citealt{Price2012a})
code {\sc phantom} \citep{PF2010,LP2010}.  Misaligned accretion disks
in binary systems have been modelled previously with {\sc phantom}
\citep[e.g.][]{Nixon2012,Nixonetal2013, Martinetal2014}.  The binary
and disk parameters are summarised in Table~\ref{tab}.  The equal mass
binary, with total mass $M=M_1+M_2$, has a circular orbit in the
$x$-$y$ plane with separation, $a$. We choose the accretion radius 
for particle removal from the simulation about each object to be $0.025\,a$.

Fig.~\ref{initial} shows the initially flat and circular but tilted
disk. The disk has a mass of $10^{-3}\,M$ with $10^6$ particles and is
inclined by $60^\circ$ to the binary orbital plane.  The value of the
disk mass has no dynamical significance in the calculation, since the
disk self-gravity is ignored, and is too low mass to affect the binary
orbit for the timescale simulated. The initial surface density of the
disk has a power law distribution $\Sigma \propto R^{-3/2}$ between
$R_{\rm in}=0.025\, a$ and $R_{\rm out}=0.25\,a$. The outer radius of
the disk is chosen to be the tidal truncation radius for the disk
assuming a coplanar binary \citep{Paczynski1977}. However, misaligned
disks feel a weaker binary torque and thus the outer truncation radius
can be much larger than this value (Nixon et al. in prep). We take a
locally isothermal disk with sound speed $c_{\rm s} \propto R^{-3/4}$
and $H/R=0.02$ at $R=R_{\rm out}$. This is chosen so that both
$\alpha$ and $\left<h \right>/H$ are constant over the disk
\citep{LP2007}. The \cite{SS1973} $\alpha$ parameter varies in the
small range $0.1-0.12$ over the disk (we implement the disk viscosity
in the usual manner by adapting the SPH artificial viscosity according
to the procedure described in \cite{LP2010}, using $\alpha_{\rm AV} =
1.91$ and $\beta_{\rm AV} = 2.0$). The disk is resolved with
shell-averaged smoothing length per scale height $\left<h\right> /H
\approx 0.52$.

\begin{figure*}
\includegraphics[width=8.4cm]{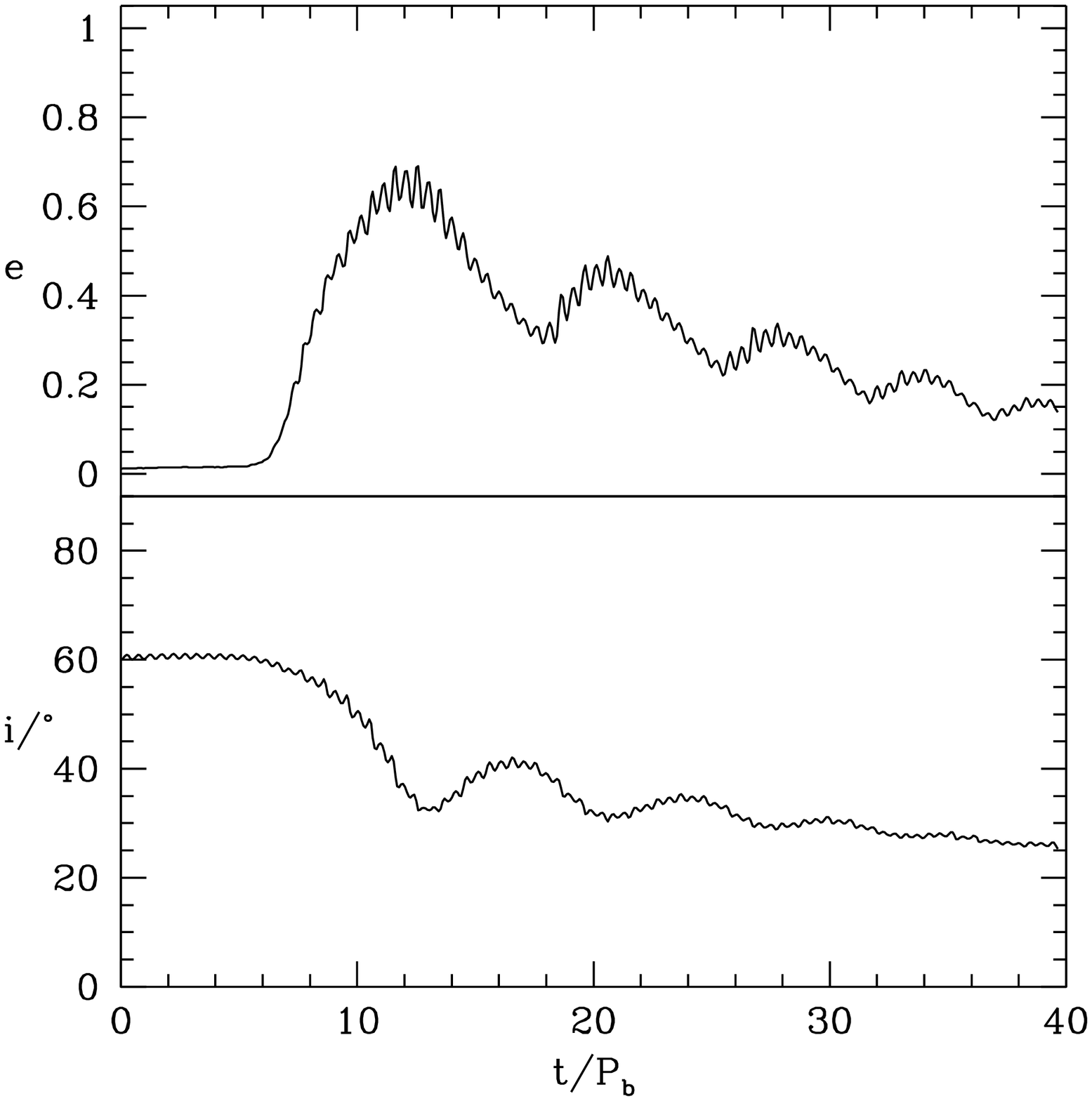}
\includegraphics[width=8.4cm]{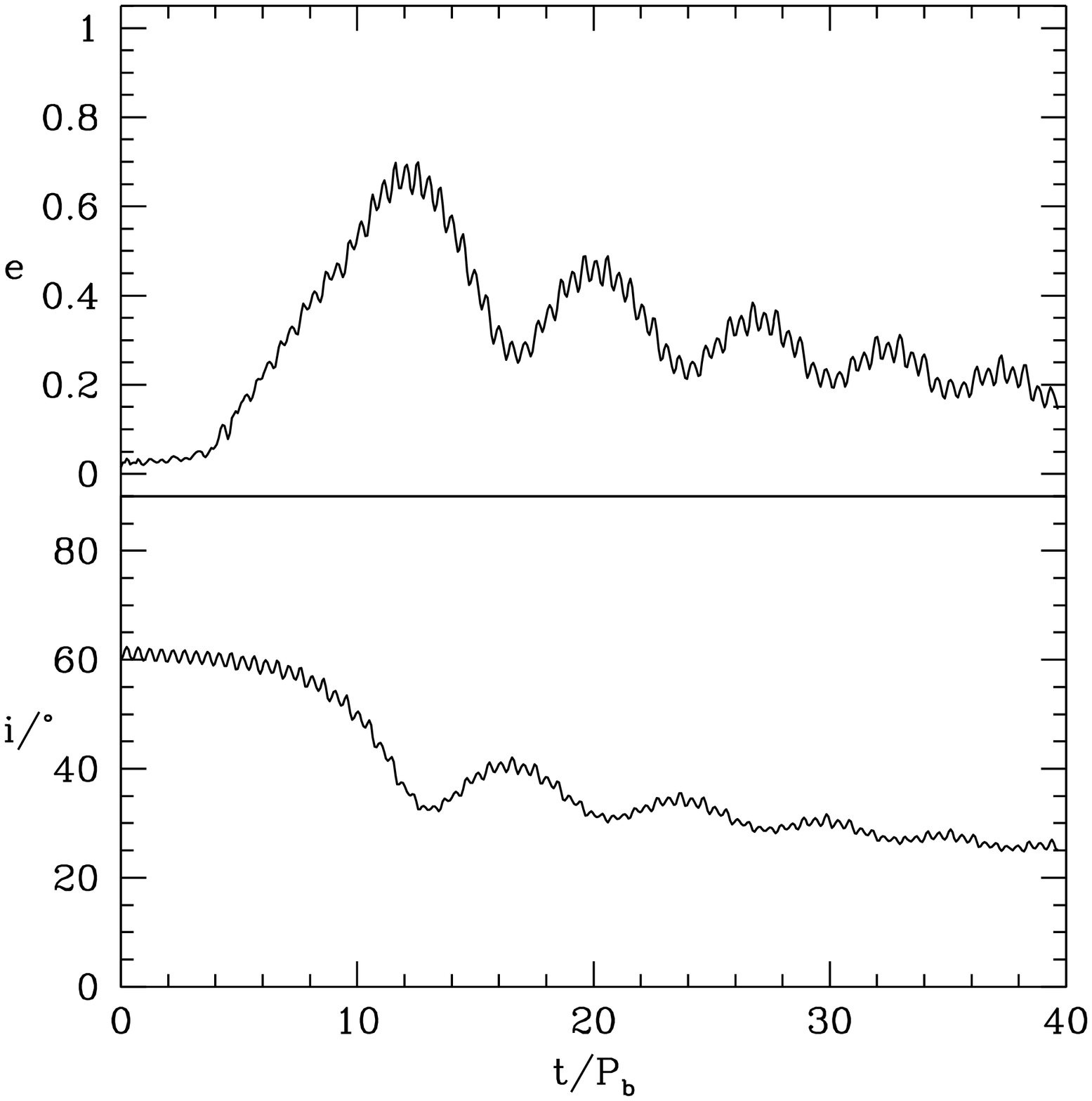}
\caption{The eccentricity and inclination evolution of the disk at a
  radius $d=0.1\,a$ (left) and $d=0.2\,a$ (right) from the
  primary.  }
\label{disk}
\end{figure*}

\begin{figure*}
\includegraphics[width=17cm]{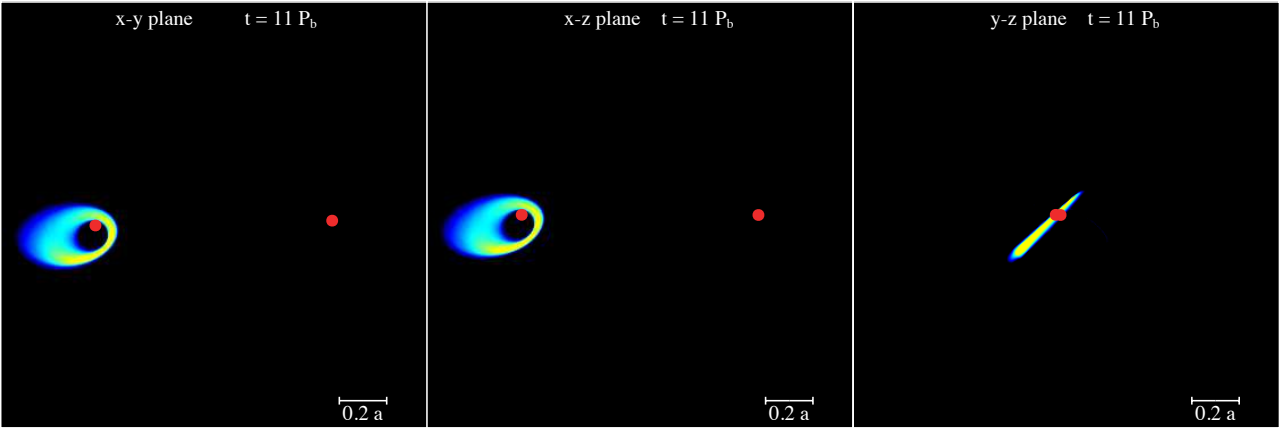}
\caption{ Same as Fig.~\ref{initial} but at the time of maximum
  eccentricity, $t=11\,P_{\rm b}$. }
\label{later}
\end{figure*}

In Fig.~\ref{disk} we show the time evolution of the eccentricity and
inclination of the disk at two radii from the primary, $d=0.1\,a$ and
$d=0.2\,a$. The figure clearly shows damped KL oscillations of the
disk. As the eccentricity increases, the inclination decreases and
vice versa. In Fig.~\ref{later} we show the disk at the maximum
eccentricity at a time of $t=11\,P_{\rm b}$. Since there is
dissipation within a disk (that is not present in a particle orbit),
the eccentricity does not reach the maximum value of $0.76$ predicted
by equation~(\ref{emax}) (see also Fig~\ref{orbits}). In addition, the
magnitude of the oscillations decays in time.  According to
equation~(\ref{tauKL}) the local oscillation timescales $\tau_{\rm
  KL}$ should differ by a factor of about 2.8 at the two radii
considered.  But we see by comparing the two plots that the evolution
of the disk at these two radii is very similar in magnitude and
timescale.  Thus, the disk is undergoing global KL
oscillations.

In our simulation the disk has a small radial extent $(R_{\rm
  out}/R_{\rm in} \approx 10)$, whereas in an astrophysical situation
the inner disk radius is often much smaller. For the simulations in
this work, the inner boundary is a circular mass sink.  If the disk
extends inward to the surface of a star, then a hard wall condition is
appropriate.  In that case, the inner boundary condition is that the
disk eccentricity vanishes \citep[see][]{Lubow2010}. Another possible
inner boundary condition is that the radial derivative of the
eccentricity vector vanishes. We discuss this further in
Section~\ref{timescales}.

The realignment of this highly misaligned disk (that is not accreting
any new material) proceeds in two stages. In the first stage, the KL
mechanism drives a rapid decay down to $i = i_{\rm cr} \approx
40^\circ$. The rapidity of the decay depends on the parameters we have
adopted and should be further explored in the future.  In the second
stage, the long term evolution of the disk proceeds due to secular
processes. In this stage, the alignment torque is due to viscous
forces that interact with a disk warp. The warp is due to binary
gravitational torques that act to precess the disk differentially,
rather than the KL torques of the first stage.  The second stage
alignment occurs on a much longer timescale than we have simulated
\citep[e.g.][]{Kingetal2013}.

%An observational study of mutual disc alignments in young binaries
%using polarization by \cite{Jensenetal2004} provides some support for
%this two stage tilt evolution. The discs in a sample of wide
%($200-1000\,\rm AU$) binaries found in Taurus-Auriga and
%Scorpius-Ophiuchus do not have random mutual orientations, as would be
%expected for their stellar spins \citep{Hale1994}.  These results
%suggest that substantial inclination evolution towards coplanarity has
%occurred. On the other hand, they typically exhibit departures from
%alignment $> 10 ^\circ$.  Since all of observed binaries have ages of
%order $10^6$ yr, even the widest systems are expected to have executed
%more than 20 binary orbits.

We have considered a range of disk parameters and we find
qualitatively the same disk behaviour. For example, when we reduce
$\alpha$ by an order of magnitude to 0.01 (more relevant to
protoplanetary disks) the KL oscillations are longer-lived due to
weaker dissipation. We have also considered a counterrotating disk
with initial inclination $i=120^\circ$.  In this case, the
eccentricity growth is similar to that in Fig.~\ref{disk}, but the
inclination of the disk evolves towards counteralignment rather than
alignment. There are a wide range of disk and binary parameters that
should be explored in future work. In the next Section we make some
order-of-magnitude estimates for binary and disk parameters for which
the KL mechanism is important.

\vspace{1cm}

\section{Disk Timescales}
\label{timescales}

In this Section we consider critical disk and binary parameters for
which the KL mechanism operates. For the KL mechanism to operate, the apsidal precession due
to the internal disk forces cannot dominate over the binary gravitational
induced apsidal rate. Otherwise, the KL effects cancel over time.

\subsection{Pressure}
The precession rate associated with pressure can be quite small
\cite[]{Lubow1992, Goodchild2006}.  The pressure induced precession
rate $|\Omega - \kappa| \sim (H/R)^2 \Omega$, where $H/R$ and $\Omega$
are evaluated at some point within the disk.  Furthermore, if the
sound crossing timescale radially across the disk is shorter than the
period of KL oscillations, the disk can communicate globally and
undergo a coherent large-scale response \citep[see e.g.][for the case
  of differential precession]{Larwoodetal1996}.  This condition can be
written as $1/\tau_{\rm KL} \la (H/R) \Omega$, while the apsidal rate
is smaller $\sim (H/R)^2 \Omega$.  For the parameters in simulation of
Figure \ref{disk}, the estimated apsidal rate is smaller than the disk
KL precession rate, and the coherence constraint is
satisfied. Consequently, the required conditions for KL global
oscillations to operate in our simulations appear to be satisfied.

\subsection{Self-Gravity}

Apsidal disk precession can also occur due to the effects of
self-gravity in the gas, but we have not included this effect in our
simulations.  We determined the local apsidal precession rate due to
self-gravity in a disk with surface density density profile $\Sigma
\propto 1/R^q$ for $q=1$ that extends from $R=0$ to $R=0.35a$ using
equations (6) and (7) of \cite{Lubow2001} with a smoothing parameter
$H$ whose square is added inside the square root of equation (5) in
that paper.
%with a smoothing parameter
%  $H$ that is added in quadrature inside the square root of equation
%  (5) in that paper.  
We identify $H$ with the disk thickness.  The (global) disk precession
rate is obtained by taking the angular-momentum weighted average of
the local rates \citep{LP1997,Lubow2001}. Based on this result, we
crudely estimate the condition for suppressing KL by self-gravity as
$M_{\rm d}\gtrsim b (H/R) M_1M_2/M$, where we estimate $b \sim 1/3$.
$b$ is fairly insensitive to $q$ for $0 < q < 1.5$.  Disks where
self-gravity is weak are relevant to a wide variety of astrophysical
scenarios such as protoplanetary, Be star, X-ray binary,
circumplanetary, and AGN disks. However, if the disk self-gravity is
sufficiently strong, as sometimes occurs in the early evolution of
disks around young stars, the KL oscillations may be suppressed
\citep{Batygin2012}.

\subsection{Global Oscillation Timescale}

Since the disk responds globally, we apply the theory of rigid disks
\citep{LP1997,Lubow2001} and estimate that the global disk
response period of inclination oscillations is
\begin{equation}
\left<\tau_{\rm KL}\right>\approx\frac{\int_{R_{\rm in}}^{R_{\rm
      out}}\Sigma R^3\sqrt{\frac{GM_1}{R^3}}\,dR }{\int_{R_{\rm
      in}}^{R_{\rm out}}\tau_{\rm KL}^{-1} \Sigma
  R^3\sqrt{\frac{GM_1}{R^3}}\,dR }.
\end{equation}
Given that equation~(\ref{tauKL}) for $\tau_{\rm KL}$ has an
inclination dependence and is therefore only valid up to a factor of a
few, this equation is also only valid up to a factor of a few. If we
assume the surface density is a power law in radius, $\Sigma \propto
R^{-p}$, then we find
\begin{equation}
\frac{\left<\tau_{\rm KL}\right>}{P_{\rm b}}\approx  \frac{(4-p)}{(5/2-p)}\frac{\sqrt{M_1M}}{M_2}\left(\frac{a}{R_{\rm out}}\right)^\frac{3}{2},
\end{equation}
for $R_{\rm out} \gg R_{\rm in}$.  With $M_1=M_2=0.5\,M$ and $p=1.5$, we
find $\left<\tau_{\rm KL}\right>=17.1(0.35 a/R_{\rm
  out})^\frac{3}{2}\,P_{\rm b}$. We have normalized the outer radius
by $0.35 a$, since the disk expands to approximately this value during
the oscillations. We find that this estimate for $\left<\tau_{\rm
  KL}\right>$ is commensurate with the timescale observed initially in
the simulation in Fig.~3 of $\tau_{\rm KL}\approx 16\,P_{\rm b}$.

\section{Discussion}

The applications of the KL disk oscillations to various astrophysical
systems depends on how these oscillations behave under different disk
conditions more generally than we have considered here. In particular,
the nature of the longevity of the oscillations needs to be
understood.  If the results found here hold generally, then the
oscillations are damped after a few dozen binary orbital periods and
the disk attains a somewhat eccentric state with inclination at the
critical value for the KL oscillations, subject to a longer timescale
decay by viscous dissipation. But the timescale for the KL decay may
not be the same for warmer disks, less viscous disks, or more extreme
mass ratio binaries.

In the case of Be/X-ray binaries, the disks are transient and the
KL oscillations may play an important role.  In \cite{Martinetal2014},
we found significant eccentricity growth in a highly misaligned
circumprimary disk of a Be star in an eccentric binary with a neutron
star companion. At the time of writing we had only investigated
eccentricity growth due to the eccentric companion in the coplanar
case. However, the results of this paper indicate that the
eccentricity growth we found for the Be star disk is due to the KL
effect.  The KL effect explains why the eccentricity growth in the
disk was only present for large inclination angles, and why its
strength did not decay with increasing tilt angle.  The eccentricity
growth is a key ingredient of our Type~II X-ray outburst model.
Thus, we would expect Type~II outbursts only in Be/X-ray binary
systems that have a Be star spin (or disk plane) misalignment in the
range $40 - 140^\circ$.

In the case of a Be star, the ratio of the outer radius of the disk to
star is not very large, around 7. In that case, an inner boundary
condition such as requiring that the eccentricity vanish will play
some role in the determining the structure of the disk eccentricity
and may have some influence on the evolution of the global structure
of the disk. 

SMBH binary disks may also be susceptible to KL oscillations. It is
often expected that SMBH binaries accrete gas from circumbinary disks
and it is possible that these circumbinary disks form highly
misaligned to the binary orbit
\citep{Nixonetal2011a,Nixonetal2011b}. It is therefore also possible
that misaligned circumprimary and circumsecondary disks can form
\citep[see e.g. Figs 6 \& 7 of ][]{Nixonetal2013} and be subject to
KL cycles. In this case, the strong increase in density at the
pericentre of the disk orbit could result in strong dissipation or
enhanced star formation.

Since binary stars are common, the disk KL mechanism presented here
could play a role in the process of planet formation around the
components of a binary.  In particular, we would like to understand
how planets that are apparently undergoing KL oscillations in wide
binaries could form in protostellar disks
\citep[e.g.][]{Wu2003,Takeda2005}. According to our calculations, the
disk tilt damps after KL oscillations to the critical angle. A planet
forming in such a disk after damping would not undergo KL
oscillations. They must have formed in a disk that avoided evolution
that damped the inclination to the critical inclination angle $i_{\rm
  cr}$ at the time of their formation.  Based on the results of this
paper, a circular binary with separation $a=10^3$ AU with each star
having $1 M_{\odot}$, would induce KL oscillations in a $\sim 350$ AU
disk with oscillation period $\sim 3 \times 10^5$ years.  The KL
decline in tilt could be avoided, for example, if the disk
self-gravity is sufficiently strong \citep{Batygin2012}. This issue
requires further study.

%We have also noted in Section \ref{sim} that there is some
%observational evidence by \cite{Jensenetal2004} that suggests some
%tilt evolution occurs during the T Tauri stage which generally brings
%the disk tilt below the threshold for KL oscillations.

\section{Conclusions}

We have found that the Kozai--Lidov mechanism, that exchanges
inclination for eccentricity in a highly misaligned particle orbit
around a component of a binary system, can also operate in a fluid
disk.  This result has implications for a range of astrophysical
systems. Much work remains to be done to understand its general
behavior. Simulations should be performed for different system
parameters. It would also be desirable to develop a linear model for
the disk evolution involving a disk whose initial inclination is just
above $i_{\rm cr}$ by extending existing linear models for
eccentricity and inclination evolution.

\section*{Acknowledgments} 
RGM's support was provided under contract with the California
Institute of Technology (Caltech) funded by NASA through the Sagan
Fellowship Program.  Support for CJN was provided by NASA through the
Einstein Fellowship Program, grant PF2-130098. SHL acknowledges
support from NASA grant NNX11AK61G. PJA acknowledges support from
NASA's ATP program under awards NNX11AE12G and NNX14AB42G. DJP is
supported by Future Fellowship FT130100034 from the Australian
Research Council. We acknowledge the use of SPLASH \citep{Price2007}
for the rendering of the figures. This work utilised the Janus
supercomputer, which is supported by the National Science Foundation
(award number CNS-0821794), the University of Colorado Boulder, the
University of Colorado Denver, and the National Center for Atmospheric
Research. The Janus supercomputer is operated by the University of
Colorado Boulder.

\bibliographystyle{apj}

\label{lastpage}
\end{document}